\documentclass[lettersize,journal]{IEEEtran}
\usepackage[utf8]{inputenc}

\usepackage{graphicx}
\ifCLASSINFOpdf
\else
\fi
\usepackage{amsfonts}
\usepackage{color}
\usepackage{soul}
\usepackage{subfigure}
\usepackage{tabularx,booktabs}
\usepackage{amssymb,amsfonts,amsthm}
\usepackage{algorithmic}
\usepackage{graphicx}
\usepackage{textcomp}
\usepackage{xcolor}
\def\BibTeX{{\rm B\kern-.05em{\sc i\kern-.025em b}\kern-.08em
    T\kern-.1667em\lower.7ex\hbox{E}\kern-.125emX}}
\usepackage{textcomp}
\usepackage{url}
\usepackage{multirow}
\usepackage{boldline}
\usepackage{pifont}
\usepackage{amsmath}               
  {
      \theoremstyle{plain}
      \newtheorem{assumption}{Assumption}
  }

\usepackage[noadjust]{cite}

\usepackage[linesnumbered,ruled]{algorithm2e}
\usepackage{subfiles}

\theoremstyle{definition}

\title{End-to-End Latency Analysis and Optimal Block Size of Proof-of-Work Blockchain Applications}

\author{Francesc Wilhelmi, Sergio Barrachina-Mu\~noz and Paolo Dini}
\begin{document}
\maketitle

\begin{abstract}
Due to the increasing interest in blockchain technology for fostering secure, auditable, decentralized applications, a set of challenges associated with this technology need to be addressed. In this letter, we focus on the delay associated with Proof-of-Work (PoW)-based blockchain networks, whereby participants validate the new information to be appended to a distributed ledger via consensus to confirm transactions. We propose a novel end-to-end latency model based on batch-service queuing theory that characterizes timers and forks for the first time. Furthermore, we derive an estimation of optimum block size analytically. Endorsed by simulation results, we show that the optimal block size approximation is a consistent method that leads to close-to-optimal performance by significantly reducing the overheads associated with blockchain applications. 
\end{abstract}


\IEEEpeerreviewmaketitle

\section{Introduction}

\IEEEPARstart{B}{lockchain} has emerged as a groundbreaking approach to enable decentralized solutions with security, trust, and resilience. Its transparency enables collaboration among several untrusted partners, so its usage has recently spanned to multiple domains, including e-health, finance, governance, and communications~\cite{al2019blockchain}, and paved the way for unprecedented collaborative applications such as joint optimization via Federated Learning (FL)~\cite{nguyen2021federated}. Nevertheless, the performance of blockchain systems raises concerns, especially for public blockchains, where anyone is allowed to participate~\cite{vukolic2015quest,swan2015blockchain}. For instance, the Proof-of-Work (PoW) consensus mechanism is well known to require a high delay between mined blocks and to incur high energy consumption.

Several works attempt to optimize blockchain systems by addressing the trade-offs between decentralization, security, and performance. This letter contributes to such efforts by tackling the minimization of the transaction confirmation latency in PoW-based blockchains. In this respect, the work in~\cite{kim2019blockchained} provided an end-to-end latency evaluation for blockchain-enabled FL to optimize the block generation rate, $\lambda$. A similar analysis was performed in~\cite{pokhrel2020federated}, where the optimal block size was obtained by minimizing the fork probability, also in the domain of FL. Another relevant work is~\cite{feng2021blockchain}, which analyzed the transaction confirmation latency and derived the optimal block generation ratio through a genetic algorithm. Alternatively,~\cite{lu2020low} studied the communication cost of blockchain on FL optimization and provided a deep reinforcement learning (DRL) mechanism to improve the system utility, which captures the trade-off between the learning time and the learning accuracy. DRL was also used in~\cite{liu2019performance} to improve the performance of a blockchain by selecting the best set of block producers, consensus algorithm, block size, and block interval.


Different from the related work, we focus on block size optimization to minimize the transaction confirmation latency in the blockchain. This problem substantially differs from the optimization of the block generation rate, which depends on the computational capacity of miners and the difficulty of the consensus problem to be solved. Instead, controlling the block size allows optimizing the blockchain system against the number of concurrent users submitting transactions. Moreover, our model includes the effect of timers and forks, which are important aspects of a PoW-based blockchain.

Our findings reveal that there exist several trade-offs to be taken into account when adjusting the block size. First, for small block sizes, the fork probability decreases (the block propagation time is lower), but more overhead is incurred to the overall transaction confirmation latency. In contrast, a large block size contributes to decreasing the overhead of the blockchain, thus potentially reducing the time transactions spend in the pool before being included in a candidate block. As we argue in this letter, finding the optimal block size analytically is unfeasible. Nevertheless, our results show that the block size approximation proposed here is a consistent method to lead to close-to-optimum performance.

\section{PoW-based Blockchain Operation}


A blockchain system is composed of a set of miners forming a peer-to-peer (P2P) network. Miners are responsible for maintaining, updating, and verifying the status of a distributed ledger, which contains the application information (e.g., economic transactions or infrastructure ownership status) generated by a set of client devices. The information stored in a blockchain is organized in cryptographically chained blocks, a principal requirement for ensuring the immutability and inviolability of the system. Each block is connected to its predecessor (until the initial genesis block) following a hash function, so any alteration on any block would alter the entire chain, which would not be accepted by the rest of the miners. While many functions exist for encoding blockchain data and chaining blocks, the most popular ones are based on Merkle tree~\cite{merkle1987digital}. Following a Merkle tree structure, a block is composed of the nonce (the result of mining), the previous block hash, the Merkle root (a composition of hashes from previous blocks), the timestamp, and a body with transactions.

\begin{figure}[ht!]
\centering
\includegraphics[width=\linewidth]{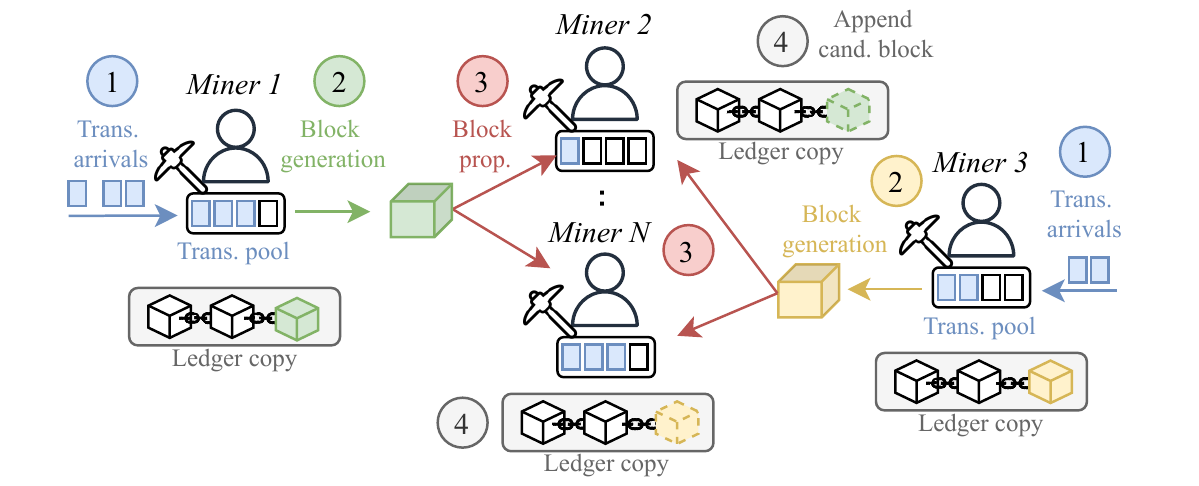}
\caption{PoW-based blockchain operation: (1) miners receive transactions from users and other miners, (2) candidate blocks are mined running consensus, (3) mined blocks are propagated, (4) miners append candidate blocks to their ledger copy (forks may occur, leading to inconsistencies).}
\label{fig:bc_overview}
\end{figure}

Blocks of variable size can be formed by gathering transactions from clients so that the block size is set as $b = h + n t$, where $h$ is the header size (constant), $n$ is the number of transactions included in the block, and $t$ is the size of a single transaction. To generate a new block, miners first gather and propagate transactions from clients either until a candidate block has enough transactions (determined by the block size $b$), or until a maximum waiting timer $\tau$ expires. Then, following the PoW consensus mechanism and using a mining capacity of $\lambda$, miners solve a computation-intensive operation. While PoW entails a significant redundancy in computation and storage, it grants a high level of security to fully decentralized environments. Fig.~\ref{fig:bc_overview} illustrates the abovementioned PoW-based blockchain operation. As shown, miners 1 and 3 gather transactions and generate a block simultaneously, which is accepted by different miners in the network (miners 2 and $N$), thus potentially leading to inconsistencies (forks).

\section{Latency Model and Block Size Optimization}


\subsection{Transaction confirmation latency}
\label{sec:e2e_latency}

The transaction confirmation latency of a blockchain can be derived by considering the following delays:
\begin{enumerate}
    \item \textbf{Queuing delay ($\text{T}_\text{q}$):} Clients submit transactions to the closest miner following a Poisson process where the time between transactions is given by an exponential distribution with parameter $\mu$. The submitted transactions wait in a pool before being included in a candidate block, i.e., when the block size $b$ is achieved or when the waiting timer $\tau$ expires. For the sake of simplicity, we assume that the transactions pool is perfectly shared. We resort to the queue model of Sect.~\ref{sec:queue_model} to characterize the queuing delay.
    \item \textbf{Block generation delay ($\text{T}_\text{bg}$):} Miners run PoW to find the candidate block's nonce. This entails solving a computation-intensive mathematical puzzle. This process is modeled through an exponential random variable with parameter $\lambda$, so the expected mining time $\text{T}_\text{bg}$ for $M$ miners is given by $(M\lambda)^{-1}$. Notice that the block size does not affect the mining time.
    \item \textbf{Block propagation delay ($\text{T}_\text{bp}$):} Mined blocks are propagated throughout the P2P network. The block propagation delay depends on the size of the P2P network and the links between miners. Blocks arrive simultaneously to all the miners to avoid synchronization issues.
\end{enumerate}

Considering the effect of forks on the end-to-end latency, we define the blockchain transaction confirmation $\text{T}_\text{BC}$ latency:
\begin{equation}
 \text{T}_\text{BC} = \frac{\text{T}_\text{q} + \text{T}_\text{bg} + \text{T}_\text{bp}}{1-p_\text{fork}},
\end{equation}
where the fork probability $p_\text{fork}$ denotes the cases when two or more miners succeed in mining a block before the winner's one is completely propagated. With forks, uninformed miners may mistakenly add non-winner blocks to their ledger version. Since the time between blocks is a Poisson inter-arrival process, the fork probability is given by $p_\text{fork} = 1 - e^{-\lambda (M-1)\text{T}_\text{bp}}$.


\subsection{Queue Model}
\label{sec:queue_model}

\begin{figure}[ht!]
\centering
\includegraphics[width=\linewidth]{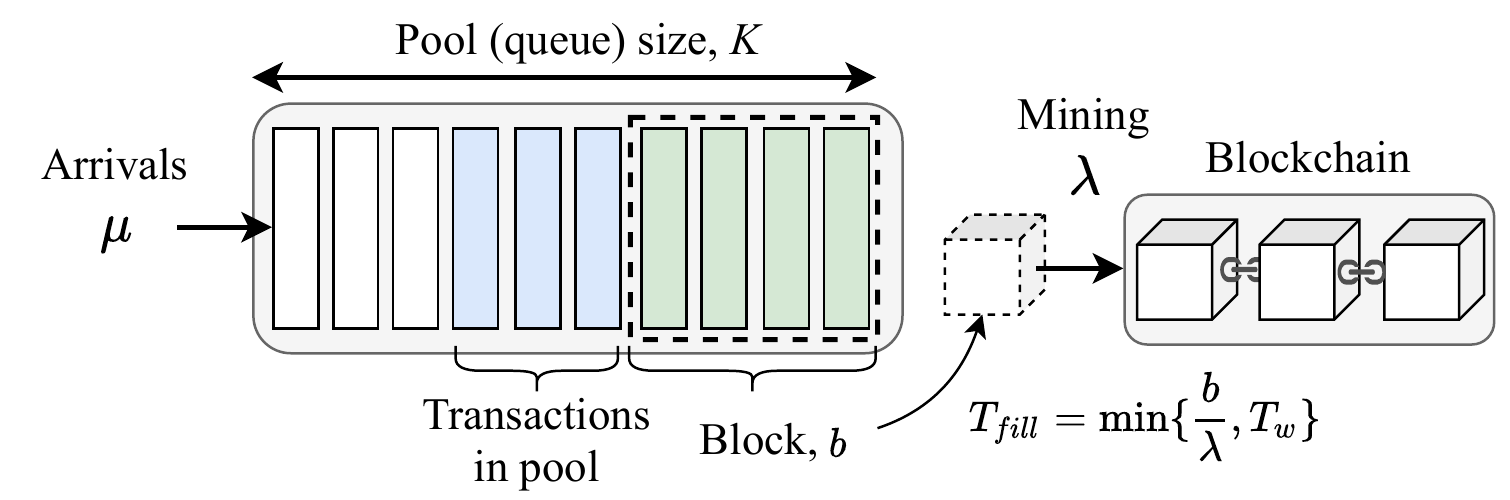}
\caption{Batch-service queue for blockchain.}
\label{fig:batch_service_queue}
\end{figure}

As done in~\cite{wilhelmi2021discrete}, we consider a finite-length $M/M^s/1/K$ queue (see Fig.~\ref{fig:batch_service_queue}) -- where $s$ denotes the number of transactions to be served in a batch -- to derive the transaction confirmation latency in blockchain. To that purpose, we model the queue using a Markov chain where states indicate the number of queued transactions before a departure. We apply the Poisson arrivals see time averages (PASTA) property to obtain the steady-state queue distribution.

To calculate the expected queue delay, we focus on the queue's steady-state distribution $\pi^k$, which indicates the probability of each state $k$:
\begin{equation}
\text{T}_\text{q} = \frac{\sum_{k=0}^K k \pi^k}{\mu(1-\pi^K)}.
\end{equation}

To get $\pi^k$, we first derive the departures distribution $\boldsymbol{\pi}^d$, which is obtained by solving $\boldsymbol{\pi}^d = \boldsymbol{\pi}^d \boldsymbol{\text{P}}$, being $\text{P}$ the transition-probability matrix and $\boldsymbol{\pi}^d \boldsymbol{1}^T = 1$:


\begin{equation}
\resizebox{.92\columnwidth}{!} 
{
    \small
    \centering
    \text{P} = \bordermatrix{& 0 & 1 & \ldots & b & \ldots & K-b & \ldots & K \cr
    \;\;\;0 & p_{0,0} & p_{0,1} & \ldots & p_{0,b} & \ldots & p_{0,K-b} & \ldots & \overline{p}_{0,K} \cr
    \;\;\;1 & p_{1,0} & p_{1,1} & \ldots & p_{1,b} & \ldots & p_{1,K-b} & \ldots & 0\cr
    \;\;\;\vdots & \vdots & \vdots & \ddots & \vdots & \ddots & \vdots & \ddots & \vdots \cr
    \;\;\;b & p_{b,0}  &  p_{b,1} & \ldots & {p}_{b,b} & \ldots & \overline{p}_{b,K-b} & \ldots & 0 \cr
    \;\;\;\:\vdots & \vdots & \vdots & \ddots & \vdots & \ddots & \vdots & \ddots & \vdots \cr
    K-b & 0  &  0 & \ldots & p_{K-b,b} & \ldots & \overline{p}_{b+1,b+1} & \ldots & 0 \cr
    \;\;\;\:\vdots & \vdots & \vdots & \ddots & \vdots & \ddots & \vdots & \ddots & \vdots \cr
    \;\;\;K & 0 & 0 & \ldots & 1 & \ldots & 0 & \ldots & 0}
    \nonumber
}
\end{equation}

Being $s(i)$ the number of served transactions from departure state $i$, transition probabilities $p_{i,j}$ and $\overline{p}_{i,j}$ are computed as


\begin{equation}
    \resizebox{0.9\columnwidth}{!}{%
        $p_{i,j} = \frac{\lambda}{\lambda + \mu}\Big(\frac{\mu}{\lambda + \mu}\Big)^{j-\big(i-s(i)\big)},     \overline{p}_{i,j} = 1-\sum^{K-s(i)-1}_{l=0} p_{i,l}$%
        }
\end{equation}

Forks are captured within the served transactions at departures. Considering that $\mathcal{T}_f$ represents the set of non-conflicting transactions that remain valid even in the event of a fork (i.e., transactions included in the winner's block and not included in other forked blocks), $s(i)$ is obtained as follows:
\begin{equation}
    s(i) = (1-p_\text{fork})\cdot\min\{i,b\} + p_\text{fork}\cdot |\mathcal{T}_{f}|.
\end{equation}

Finally, the steady-state queue distribution $\boldsymbol{\pi}^k$ for $k < K$ is given by
\begin{equation}    \label{eq:steady}
\begin{split}
\pi^k &= \frac{1}{\mu \text{T}_\text{d}} \sum_{i=0}^k \pi_i^d \bigg[ \overline{\varsigma}^{\tau}_{i} \Big( \sum_{j=k-s(i)+1}^{K-s(i)} p_{i,j} \Big) \\
&+ \varsigma^{\tau}_{i} \Big( \sum_{j=i}^{b-1} \big(\text{Pr}(j-i|\tau) \sum_{l=k-s(j-i)+1}^{K-s(j-i)} p_{j,l}\big) \Big) \bigg],
\end{split}
\end{equation}
where $\text{T}_\text{d}$ is the expected inter-departure time, $\varsigma^{\tau}_i$ is the timer expiration probability from departure state $i$, and $\text{Pr}(n|\tau)$ is the probability that $n$ packets arrive to the queue before the timer expires. The 
probability of finding the queue full is calculated as $\pi^K = 1-\sum^{K-1}_{k=0} \pi^{k}$.

\subsection{Optimal block size}
We now derive the optimal block size $b^*$ that minimizes the transaction confirmation time in a blockchain system. To that purpose, we use the latency model provided in Sect.~\ref{sec:e2e_latency} together with the following assumption.
 



\begin{assumption}
\label{as:3}
The waiting timeout is set to an arbitrarily large number so that the number of mined transactions is always set to the block size, $b$.
\end{assumption}

Assumption~\ref{as:3} is necessary to find a convex function that is a close fit to data obtained from the model. The fact is that the complex interactions at the queue prevent finding a convex function that allows optimizing the queuing delay. Furthermore, we approximate $\text{T}_\text{q}$ using the Lagrange interpolation method~\cite{abramowitz1970handbook} so that the queuing delay is approximated by an $n$-order polynomial from a set of $N+1$ data points obtained from the queuing model. Assuming fixed $\mu$ and $\lambda$ parameters (provided by the scenario), we approximate $T_\text{q}$ as:

\begin{equation}
    \text{T}_\text{q}(b) \approx \hat{\text{T}}_\text{q}(b) = \sum_{j=0}^{N} y_j \mathcal{L}_{N,j}(b),
\end{equation}
where $y_j$ corresponds to the queuing delays obtained from the model, and $\mathcal{L}_{N,j}(b)$ are the Lagrange basis polynomials, defined as $\prod_{k=0,k\neq j}^{N} \frac{b-b_k}{b_j-b_k}$. Being $C\in \mathbb{R}^+$ a positive constant capturing the P2P end-to-end capacity, the overall transaction confirmation delay, $\text{T}_\text{bp}$, is approximated as follows:

\begin{equation}
\hat{\text{T}}_\text{BC} = \frac{\hat{\text{T}}_\text{q}(b) + 1/M\lambda + b/C}{e^{-\lambda(M-1)b/C}}
\label{eq:tbc}
\end{equation}

Following Eq.~\eqref{eq:tbc}, which is convex for $b \geq 0$, the optimal block size $b^*$ can be derived directly. 

\section{Simulation Results and Validation}

In this Section, we assess the accuracy of the proposed block size optimization model. We include cases where Assumption~\ref{as:3} above does not hold (i.e., with realistic timer and fork behaviors), which allows comparing the performance of the actual optimal block size with its approximation by our proposed optimization. We also compare the latency model with simulations~\cite{batchsim} (see Table~\ref{tab:sim_parameters}).  

\begin{table}[ht!]
\centering
\caption{Model/Simulation parameters}
\label{tab:sim_parameters}
\begin{tabular}{@{}ccc@{}}
\toprule
Parameter & Description & Value \\ \midrule
$b$ & block size & 1-10 trans.\\
$t$ & transaction length & 5 kbits \\
$h$ & block header length & 20 kbits \\ 
$K$ & queue size & 10 trans. \\
$M$ & number of miners & \{1,10\} \\
$C$ & P2P links' capacity & 5 Mbps \\
$t_s$ & sim. time & 100,000 s \\ 
\hline
$\tau$ & mining timer & \{0.1, 1, 5, 10, 100\} s \\ 
$\mu$ & packet arrivals & \{0.1, 0.25, 0.5, 1, 2.5, 5, 10\} tps \\ 
$\lambda$ & mining rate & \{0.1, 0.25, 0.5, 1, 2.5, 5, 10\} Hz \\
\bottomrule
\end{tabular}
\end{table}

\begin{figure}[ht!]
\centering
\includegraphics[width=\linewidth]{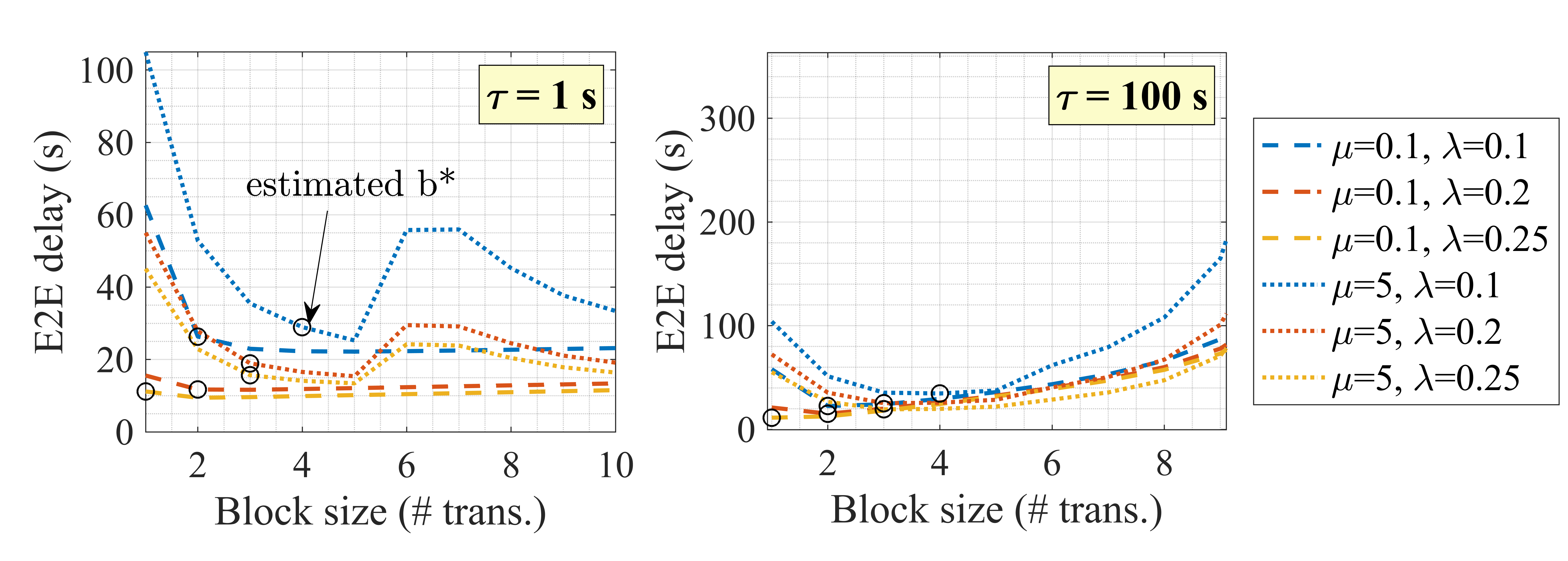}
\caption{Transaction confirmation latency for different BC parameters: (a) $\tau = 1$ s, (b) $\tau = 100$ s. The delay values obtained from the estimated optimal block size ($b^*$) are represented by the black circles.}
\label{fig:timer_vs_no_timer}
\end{figure}

Fig.~\ref{fig:timer_vs_no_timer} shows the transaction confirmation latency against the block size and highlight the range of validity of the optimal block size estimation. In particular, we have considered $\mu = \{0.1, 0.25, 5\}$ arrivals per second, $\lambda = \{0.1, 0.2, 0.25\}$ Hz, $\tau = \{1, 100\}$ seconds, and disabled forks. While using $\tau=100$ s provides a smooth function whereby the optimal block size is derived, a small timer such as $\tau=1$ s breaks the convexity properties of the delay function.

As illustrated by Fig.~\ref{fig:timer_vs_no_timer}, the estimated optimal block size (highlighted with circles) matches the best result in every setting when the timer is $\tau = 100$ s (right figure). In this case, Assumption~\ref{as:3} is fulfilled due to the high timer value (blocks are mined when reaching the established-set block size). In contrast, with tighter timers (left figure), the real optimum is slightly different from the approximated one. Nevertheless, the approximation serves as a consistent heuristic and leads to a low (near-optimal) end-to-end latency.

\begin{figure}[ht!]
\centering
\includegraphics[width=1\linewidth]{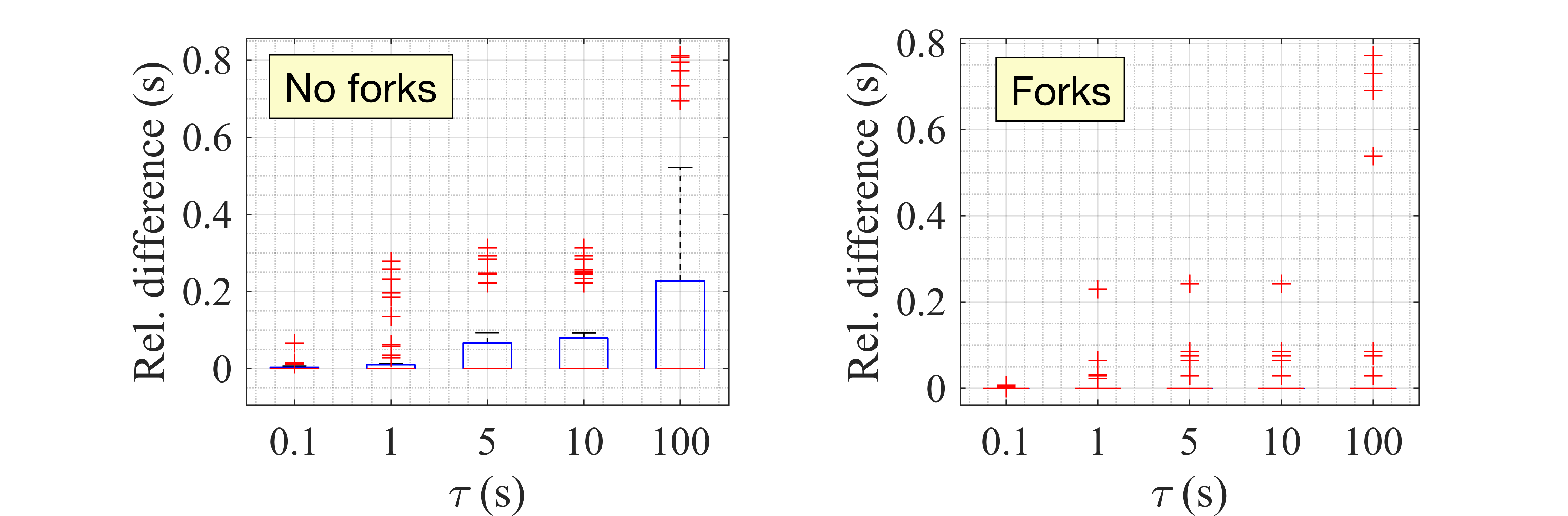}
\caption{Relative difference between the total delay obtained using $b^*$ and $\hat{b}^*$: (a) without forks ($M=1$), (b) with forks ($M=10$).}
\label{fig:approximation}
\end{figure}

\begin{figure}[ht!]
\centering
\subfigure[Without forks ($M=1$)]{\includegraphics[width=\columnwidth]{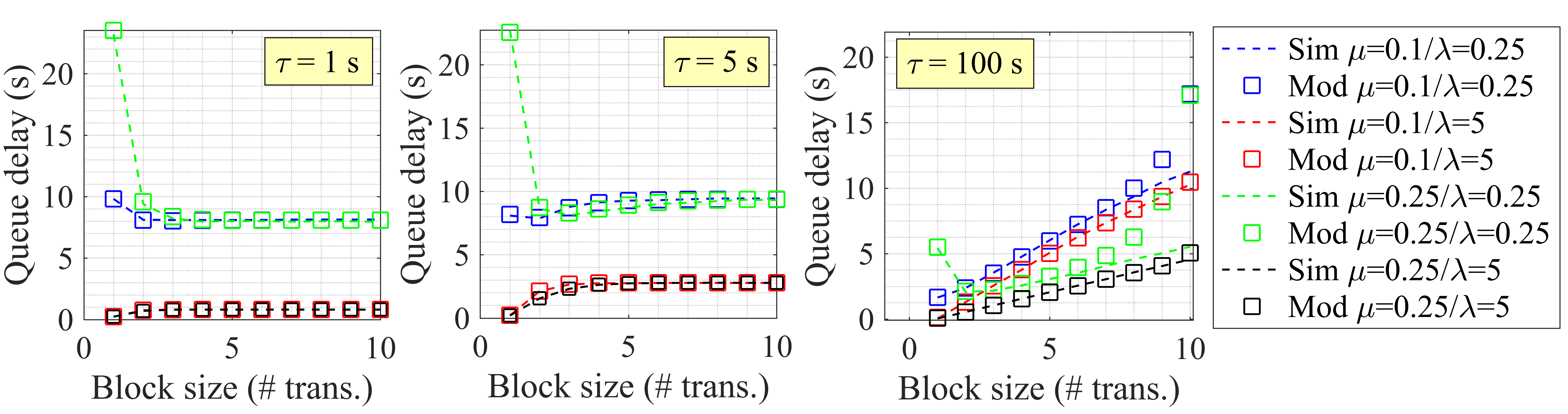}\label{fig:validation_no_forks}} 
\subfigure[With forks ($M=10$)]{\includegraphics[width=\columnwidth]{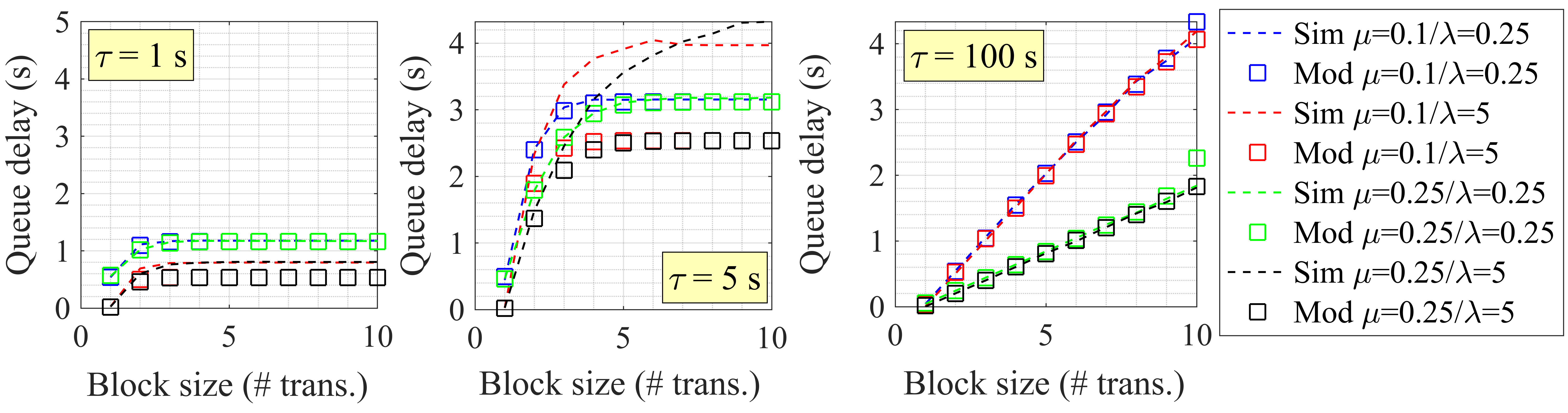}\label{fig:validation_forks}} 
\caption{Blockchain queuing delay for different $\tau$, $\mu$, $\lambda$, and $b$ parameters: (a) without forks ($M=1$), (b) with forks ($M=10$).}
\label{fig:validation}
\end{figure}

To reinforce this statement, Fig.~\ref{fig:approximation} compares the performance obtained by the actual optimal block size (computed by brute force through simulations) and its model approximation. As shown in Fig.~\ref{fig:approximation}, the major gap is found when timers are set to high values and forks are not possible (because the number of miners is $M=1$). As shown later in Fig.~\ref{fig:validation}, this occurs for high values of $\mu$. Instead, when we consider multiple miners ($M=10$), low difference between the estimated delay and the actual minimum is noticed. Despite the convex transformation of the model function may fail to characterize the actual queue latency exactly, the estimated trend is the same, so the optimal block size can be estimated in most of the cases.


To conclude, we compare the model and simulator outputs on the blockchain queuing delay. For this analysis, we consider $\mu = \{0.1, 0.25\}$ arrivals per second, $\lambda = \{0.25, 5\}$ Hz, $\tau = \{1, 5, 100\}$ seconds, and $b = [1-10]$ transactions. The results are illustrated in Fig~\ref{fig:validation}, where model and simulation output match except for certain special cases. 

Without forks (Fig.~\ref{fig:validation_no_forks}), the model deviates from simulation when the queue is saturated with transactions ($\mu=0.25$ and $\lambda=0.25$, represented in green). For high block sizes, the model becomes unstable, given the high amount of incoming traffic. Notice that simulations consider a finite amount of time, so the transitory phase before the queue is full is also captured. Moreover, the model and the simulator results diverge when forks are possible and the waiting timer is low (i.e., $\tau = 1$ s and $\tau = 5$ s). This case confirms that the effects caused by forks cannot be fully captured by our model. The fork probability varies depending on the state from which a departure occurs, so the higher the number of transactions to be mined, the higher the fork probability. However, to capture this, it is required to have a-priori knowledge of the queue's steady-state distribution, which is in fact the output of the model. Yet, the fork probability is estimated by the model based on the input block size $b$, which is assumed to be fixed. This, however, does not hold when timers are small, provided that smaller blocks can be mined even if not enough transactions arrive before the timer expires. Consequently, it leads to a lower fork probability in practice, given that the expected mining size is lower. This effect is exacerbated as the timer and number of arrivals decrease.

\section{Conclusions}
\label{section:conclusions}



To consolidate the adoption of blockchain in practical settings, in this letter, we have addressed the optimization of the block size, an important parameter to determine the amount of information to be included in a single block. Adjusting the block size is essential to minimize the latency of PoW-based blockchain applications. However, the complex influence of timers and forks on the queuing delay prevents deriving the optimal block size in closed-form. Nevertheless, we have shown that an approximation of the queue delay function allows estimating the optimal block size, thus leading to near-optimal delay estimations. As future work, we envision the improvement of the queue model through tailored data-driven approaches.


\section*{Acknowledgment}


This work has been partially funded by the IN CERCA grant from the Secretaria d'Universitats i Recerca del departament d'Empresa i Coneixement de la Generalitat de Catalunya, by the Spanish grant PID2020-113832RB-C22(ORIGIN)/AEI/10.13039/501100011033, and by the European Union’s Horizon 2020 research and innovation programmes under Grants Agreements No. 953775 (GREENEDGE) and No. 101017171 (MARSAL).
	
\ifCLASSOPTIONcaptionsoff
\newpage
\fi

\bibliographystyle{IEEEtran}
\bibliography{bibliography}

\begin{thebibliography}{10}
\providecommand{\url}[1]{#1}
\csname url@samestyle\endcsname
\providecommand{\newblock}{\relax}
\providecommand{\bibinfo}[2]{#2}
\providecommand{\BIBentrySTDinterwordspacing}{\spaceskip=0pt\relax}
\providecommand{\BIBentryALTinterwordstretchfactor}{4}
\providecommand{\BIBentryALTinterwordspacing}{\spaceskip=\fontdimen2\font plus
\BIBentryALTinterwordstretchfactor\fontdimen3\font minus
  \fontdimen4\font\relax}
\providecommand{\BIBforeignlanguage}[2]{{%
\expandafter\ifx\csname l@#1\endcsname\relax
\typeout{** WARNING: IEEEtran.bst: No hyphenation pattern has been}%
\typeout{** loaded for the language `#1'. Using the pattern for}%
\typeout{** the default language instead.}%
\else
\language=\csname l@#1\endcsname
\fi
#2}}
\providecommand{\BIBdecl}{\relax}
\BIBdecl

\bibitem{al2019blockchain}
J.~Al-Jaroodi and N.~Mohamed, ``Blockchain in industries: A survey,''
  \emph{IEEE Access}, vol.~7, pp. 36\,500--36\,515, 2019.

\bibitem{nguyen2021federated}
D.~C. Nguyen, M.~Ding, Q.-V. Pham, P.~N. Pathirana, L.~B. Le, A.~Seneviratne,
  J.~Li, D.~Niyato, and H.~V. Poor, ``{Federated learning meets blockchain in
  edge computing: Opportunities and challenges},'' \emph{IEEE Internet of
  Things Journal}, 2021.

\bibitem{vukolic2015quest}
M.~Vukoli{\'c}, ``{The quest for scalable blockchain fabric: Proof-of-work vs.
  BFT replication},'' in \emph{International workshop on open problems in
  network security}.\hskip 1em plus 0.5em minus 0.4em\relax Springer, 2015, pp.
  112--125.

\bibitem{swan2015blockchain}
M.~Swan, \emph{``Blockchain: Blueprint for a new economy''}.\hskip 1em plus
  0.5em minus 0.4em\relax O'Reilly Media, Inc., 2015.

\bibitem{kim2019blockchained}
H.~Kim, J.~Park, M.~Bennis, and S.-L. Kim, ``Blockchained on-device federated
  learning,'' \emph{IEEE Communications Letters}, vol.~24, no.~6, pp.
  1279--1283, 2019.

\bibitem{pokhrel2020federated}
S.~R. Pokhrel and J.~Choi, ``Federated learning with blockchain for autonomous
  vehicles: Analysis and design challenges,'' \emph{IEEE Transactions on
  Communications}, vol.~68, no.~8, pp. 4734--4746, 2020.

\bibitem{feng2021blockchain}
L.~Feng, Y.~Zhao, S.~Guo, X.~Qiu, W.~Li, and P.~Yu, ``Blockchain-based
  asynchronous federated learning for internet of things,'' \emph{IEEE
  Transactions on Computers}, 2021.

\bibitem{lu2020low}
Y.~Lu, X.~Huang, K.~Zhang, S.~Maharjan, and Y.~Zhang, ``Low-latency federated
  learning and blockchain for edge association in digital twin empowered 6g
  networks,'' \emph{IEEE Transactions on Industrial Informatics}, vol.~17,
  no.~7, pp. 5098--5107, 2020.

\bibitem{liu2019performance}
M.~Liu, F.~R. Yu, Y.~Teng, V.~C. Leung, and M.~Song, ``{Performance
  optimization for blockchain-enabled industrial Internet of Things (IIoT)
  systems: A deep reinforcement learning approach},'' \emph{IEEE Transactions
  on Industrial Informatics}, vol.~15, no.~6, pp. 3559--3570, 2019.

\bibitem{merkle1987digital}
R.~C. Merkle, ``A digital signature based on a conventional encryption
  function,'' in \emph{Conference on the theory and application of
  cryptographic techniques}.\hskip 1em plus 0.5em minus 0.4em\relax Springer,
  1987, pp. 369--378.

\bibitem{wilhelmi2021discrete}
F.~Wilhelmi and L.~Giupponi, ``Discrete-time analysis of wireless blockchain
  networks,'' in \emph{2021 IEEE 32nd Annual International Symposium on
  Personal, Indoor and Mobile Radio Communications (PIMRC)}.\hskip 1em plus
  0.5em minus 0.4em\relax IEEE, 2021, pp. 1011--1017.

\bibitem{abramowitz1970handbook}
M.~Abramowitz and I.~A. Stegun, \emph{Handbook of mathematical functions with
  formulas, graphs, and mathematical tables}.\hskip 1em plus 0.5em minus
  0.4em\relax US Government printing office, 1970, vol.~55.

\bibitem{batchsim}
F.~Wilhelmi and L.~Giupponi, ``{Blockchain-oriented Batch Service Queue
  Simulator},'' \url{https://doi.org/10.5281/zenodo.4680438}, 2021.

\end{thebibliography}

\end{document}